\documentclass[prl,twocolumn,showpacs,superscriptaddress,aps]{revtex4-1}
\usepackage{graphicx,amssymb,amsmath,psfrag,color, dsfont, braket, gensymb}
\usepackage[colorlinks=true,citecolor=blue,linkcolor=blue,urlcolor=blue]{hyperref}
\begin{document}
\definecolor{darkgreen}{rgb}{0,0.5,0}
\newcommand{\mz}[1]{{\color{red} MZ: #1}}
\newcommand{\fp}[1]{{\color{blue} FP: #1}}
\newcommand{\mpst}{MPS$^{2\,}$}
\newcommand{\tebdsq}{TEBD$^{2\,}$}
\newcommand{\revision}[1]{{\color{blue} #1}}

\title{Isometric tensor network states in two dimensions}
\bibliographystyle{apsrev_nurl}

\author{Michael P. Zaletel}
\affiliation{Department of Physics, University of California, Berkeley, CA 94720, USA}
\author{Frank Pollmann}
\affiliation{Technische Universit{\"a}t M{\"u}nchen, Physics Department T42, 85747 Garching, Germany}
\affiliation{Munich Center for Quantum Science and Technology (MCQST), Schellingstr. 4, D-80799 M\"unchen}
\date{\today}

\begin{abstract}
Tensor network states (TNS) are a promising but numerically challenging tool for simulating two-dimensional (2D) quantum many-body problems. We introduce an isometric restriction of the TNS ansatz  that allows for highly efficient contraction of the network. We consider two concrete applications using this  ansatz. First, we show that a matrix-product state representation of a 2D quantum state can be iteratively transformed into an isometric 2D TNS. Second, we introduce a 2D version of the time-evolving block decimation algorithm (\tebdsq) for approximating of the ground state of a Hamiltonian as an isometric TNS---which we demonstrate for the 2D transverse field Ising model.
\end{abstract}

\maketitle

%
Overcoming the exponential growth of complexity when simulating quantum many-body systems is one of the most challenging  goals in computational physics.
For ground state properties of one-dimensional systems (1D) this challenge was answered by the density matrix renormalization group (DMRG) algorithm, which provides an essentially exact numerical solution of gapped 1D lattice models \cite{White:1992} and field theories \cite{Verstraete:2010}.
Subsequently understood as a variational method over the class of matrix product states (MPS) \cite{Ostlund:1995,Dukelsky:1998} its success follows from the ability of MPS to adequately capture the area-law entanglement characteristic of gapped ground states \cite{Hastings:2007}.
A central goal  has been to generalize the success of DMRG to higher dimensions. 
For certain classes of states, this is achieved by so called tensor network states (TNS) whose connectivity reflects the geometry of many-body entanglement \cite{nishino:2001, verstraete:2004}.
However, while evaluating properties of 1D MPS is highly efficient (scaling with the tensor dimension $\chi$ and system size $N$ as $N \chi^3$),  \emph{exactly} evaluating properties of TNS in higher dimensions is generically exponentially hard.
Consequently, there has been a long-standing effort to determine the best way to numerically approximate TNS contractions in order to minimize the variational energy of TNS for given a Hamiltonian.
Progress has been made for two-dimensional (2D) systems by introducing a number of algorithms to manipulate and optimize TNS for various  lattice models \cite{Nishino1996,Nishino1997,nishino:2001,verstraete:2004,Levin2007,Jordan2008,Jiang2008,Xie2009,Xie2012,Corboz2016,Vanderstraeten2016,Banuls08, fishman2018faster,Zheng1155, Corboz2013,Corboz2013, Liao2017}.
However, at this point it is fair to say that the ``right'' way to generalize 1D DMRG is not yet agreed upon.

In this work, we study a restriction of the TNS ansatz, which we dub ``isoTNS,'' which allows for highly efficient contraction of the network.
When collapsing either the rows or columns of the 2D network it reduces to the  canonical form of a 1D MPS \cite{Vidal2003a,PerezGarcia08}.
As a result, any 1D MPS algorithm, such as DMRG \cite{White:1992} or the time dependent block decimation (TEBD) \cite{Vidal2003a}, can be turned into a 2D algorithm  by applying it in a nested loop with respect to the rows and columns of the 2D isoTNS.
While the ansatz we discuss is  known to some practitioners \cite{PrivComm}, and is related to a previous work on correlated contour states, \cite{richter1994} it does not seem to have been studied in practice.
Here we introduce a key procedure for manipulating  isoTNS,  the ``Moses Move'' (MM), and  demonstrate its utility with two concrete applications: First we show that a 1D MPS representation of a 2D quantum state can be iteratively transformed into an isoTNS, and examine the resulting entanglement properties.
Second we implement a ``\tebdsq'' algorithm and use it to approximate the ground state of the 2D transverse field Ising model as an isoTNS.

\emph{The isometric tensor network ansatz}. 
\begin{figure}
\centering
\includegraphics[width=0.99\columnwidth]{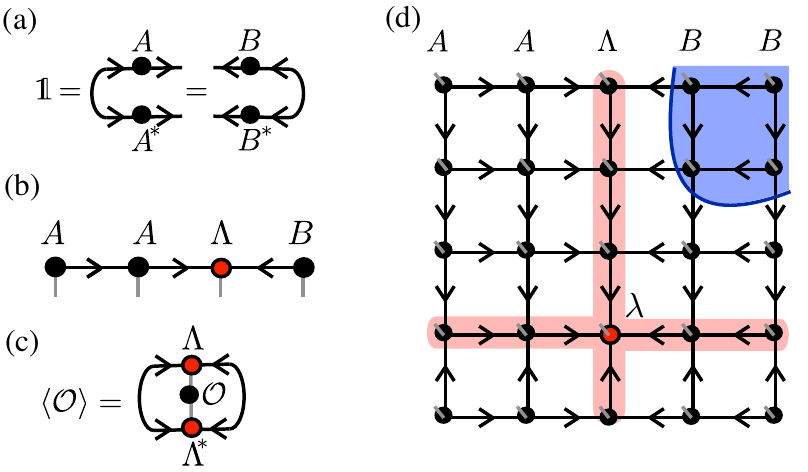}
\caption{ Schematic representation of the canonical form in 1D and 2D. 
(a) Left and right isometries are represented by arrows whose orientation indicates whether $A^{\dagger}A=BB^{\dagger}=\mathds{1}$. We view the isometry as an RG-like procedure from the large Hilbert space  (incoming arrows)  to the smaller one (outgoing arrows). 
In the case of higher-rank tensors, the contraction $A^{\dagger}A = \mathds{1}$ is always over all the incoming arrows.
(b) A 1D MPS can be brought into a mixed canonical form with orthogonality center $\Lambda$. Note that each dangling physical index implicitly has an incoming arrow.
(c) Expectation values of local operators can be directly obtained from $\Lambda$. 
(d) 2D canonical form with  ``orthogonality hypersurfaces'' $\Lambda$ (column and row highlighted in red). The orthogonality center  $\lambda$ is marked by a red dot. In blue we indicate an example of a subregion with only outgoing arrows,  whose boundary map is consequently an isometry.}
\label{fig:canon}.
\end{figure}
We first review the canonical form of a 1D  MPS (see Ref.~\cite{schollwoeck:2011} for more details). Suppressing the indices of all tensors,  the MPS for an $N$-site chain takes the form $\Psi =  T^{1} T^{2} \cdots T^{N}$. Here each  $T^{a}$ is a rank-3 tensor which we view as a $\chi_{a-1} \times \chi_{a}$ matrix in an ``ancilla space'' whose entries are vectors in the $d$-dimensional single-site Hilbert space of site $a$. Multiplication of the matrices implicitly comes with a tensor product over the single-site Hilbert spaces, producing an $N$-site wavefunction. At the boundaries, $\chi_0 = \chi_N = 1$.
For any contiguous region of spins $V = a:b$,  the partial contraction $T^{V \to \partial V} \equiv T^a \cdots T^{b}$ is a linear map from the  Hilbert space $\mathcal{H}_{V}$ of the subregion to the $\chi_{a-1} \times \chi_b$ dimensional Hilbert space $\mathcal{H}_{\partial V}$ of the ancillas  dangling from the boundary of the region. 
The ``canonical form with $\ell$-site center'' is defined by  requiring that the boundary map $T^{V \to \partial V }$ is an isometry if $V = 1$:$a$ for $a < \ell$ or $V = a$:$N$ for $a > \ell$.
Recall a map is an isometry if $T^\dagger_{V \to \partial V } T_{V \to \partial V } = \mathds{1}_{\partial V}$, while $T_{V \to \partial V } T^\dagger_{V \to \partial V } = P_{V}$ is a projection operator. The isometry condition ensures the ancillas on $\partial V$ form an orthonormal sub-basis for $V$. In what follows we denote the isometry conditions graphically by assigning arrows to the tensors as shown in Fig.~\ref{fig:canon}a \cite{Stoudenmire2013, Haegeman2016, Bal2016}.
A convenient notation for the representation of MPS with $\ell$-site center is to distinguish  the tensors $ A, \Lambda, B$ and write
\begin{align}
\Psi =  A^{1} A^{2} \cdots A^{\ell-1} \Lambda^{\ell} B^{\ell+1} \cdots B^{N}
\label{eq:1dcan}
\end{align}
as shown in Fig.~\ref{fig:canon}b.
It is easy to verify that the canonical form is satisfied if and only if each $A^a, B^a$ is individually an isometry from the left/right respectively. 
$T_{V \to \partial V }$ is an isometric boundary map if and only if the boundary $\partial V$ has only {outgoing} arrows. On the other hand, a region with only {incoming} arrows, like the $\Lambda$,  is precisely the wavefunction of the system expressed in a orthonormal basis, so it is called an \emph{orthogonality center}.
In particular, $\|\Psi\| = \|\Lambda\|$ and any site-$\ell$ expectation value can be locally computed as $\bra{\Psi} O^\ell \ket{\Psi} = \bra{\Lambda} O^\ell \ket{\Lambda} $, as seen in Fig.~\ref{fig:canon}c, because the $A, B$ tensors in its exterior contract to $\mathds{1}$ by the isometry condition.

Once the canonical form is understood as a restriction on the boundary maps, it can naturally be generalized to higher dimensions. 
By analogy to Eq.~\eqref{eq:1dcan},  we  demand that each row and column of the TNS is an isometry, as indicated in Fig.~\ref{fig:canon}d.
This constraint can be satisfied by further demanding that each tensor is an isometry from a physical and two ancilla legs to the remaining two ancillas according to the direction of the arrows indicated.
This gives a causal structure to the tensor-network, though in our convention time flows opposite to the direction of the arrows.
As in 1D, there is a set of space-like hypersurfaces with only outgoing arrows whose ``past'' defines the  wavefunction in a orthonormal basis and whose ``future'' is an isometric boundary map.
An expectation value $\bra{\Psi} O \ket{\Psi}$ depends only on the tensors in the past of the insertion $O$.
This is more than an analogy: the network between two space-like surfaces defines a Kraus decomposition of a quantum channel (time evolution) relating the boundary ancilla.

There are a special row and column $\Lambda$ (highlighted in red in Fig.~{\ref{fig:canon}}d) which has only incoming arrows, which is hence the 1D ``orthogonality hypersurface'' of the TNS (our nomenclature anticipates the generalization to higher dimensions, where $\Lambda$ has codimension one.) 
Because its exterior  is an isometry from the physical to the incoming ancillas, $\Lambda$  is the wavefunction of the system in an orthonormal basis.
Hence $\Lambda$ can be treated just like an MPS and can itself put into 1D canonical form (consequently \emph{its} orthogonality center tensor, $\lambda$, can be moved freely using the standard 1D algorithm).
Tracing over the left or right ancillas of $\Lambda$ results in a density matrix which is iso-spectral to the reduced density matrix of the right or left, e.g. $\rho_{L} \sim \Lambda \Lambda^\dagger$, so $\Lambda$ encodes the entanglement spectrum.
For any operator $O$ inside $\Lambda$,  $\bra{\Psi} O \ket{\Psi} = \bra{\Lambda} O \ket{\Lambda}$, e.g., there is a dimensional reduction to a 1D expectation value which can be computed efficiently without further approximations via standard MPS algorithms.
This is in stark contrast to generic TNS where expectation values require an approximate contraction of the entire network using, e.g., boundary MPS \cite{Verstraete2004} or corner transfer matrices \cite{Nishino1996,Nishino1997}.
Furthermore, any variationally optimal compression of the orthogonality hypersurface $\Lambda$ (such as truncation of its entanglement spectrum via SVD) is variationally optimal for the global state.
{Note that by our choice of isometries, the resulting orthogonality hypersurface $\Lambda$ has minimal entanglement for each vertical cut and is expected to follow a 1D area law. This entanglement is different from the vertical entanglement of the \emph{full} many-body wave function: it differs by the action of the isometries, which contain the 2D area law entanglement.}

The great utility of both properties will become clear in the \tebdsq algorithm we propose below.
 
It is an interesting and open question how the variational power of an isoTNS differs from that of a generic TNS.
One  restriction is that many of its correlations must decay exponentially, because any two-point function along the orthogonality hypersurface can be reduced to that of the MPS $\Lambda$, which must have exponentially decaying correlations.
In contrast, a generic 2D TNS can represent power-law correlations.
On the other hand, we have shown that any string-net state, thought to represent all 2D topological orders with gappable edges, can explicitly be put into isoTNS form.\cite{soejima2019isometric}


\emph{Shifting the orthogonality hypersurface}.
The canonical form is only useful for computational purposes if the orthogonality hypersurface $\Lambda$ can be  moved throughout the network efficiently.
In 1D,  for example, the basic move $\Lambda^\ell B^{\ell +1} = A^\ell \Lambda^{\ell + 1}$ can be accomplished by any orthogonal matrix factorization, i.e.  QR or a singular value  decomposition (SVD).
In 2D we need to solve the same equation but with $A, \Lambda, B$  entire columns of the TNS. Using QR or SVD is  hopeless, as it will destroy the locality required to express $\Lambda$ as an MPS.
The key insight is that the canonical form can be preserved under a unitary insertion  $(A^\ell U^\dagger) (U \Lambda^{\ell + 1})$.
We  propose to use this ambiguity to choose $A^\ell$ such that it ``disentangles'' $\Lambda^{\ell+1}$, so that $\Lambda^{\ell+1}$ has an efficient (low rank) MPS form.
\begin{figure}
\centering
\includegraphics[width=0.95\columnwidth]{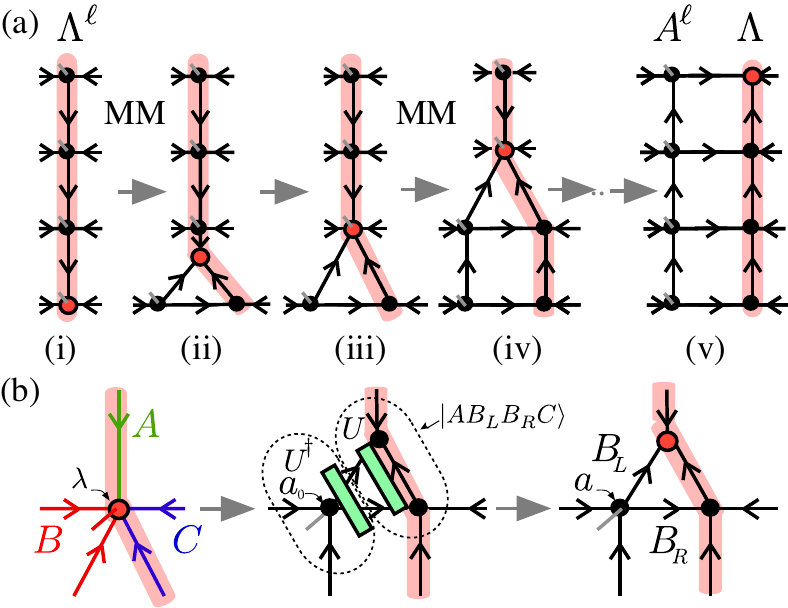}
\caption{The Moses Move. (a) The orthogonality hypersurface $\Lambda^{\ell}$ is split into the product of a left isometry $A^{\ell}$ and a zero-column state $\Lambda$ with no physical indices.
The unzipping is performed by successively applying the splitting procedure shown in panel (b).
The legs of the center site $\lambda$ are grouped into a tripartite state $\ket{ABC}$  which is ``split'' into three tensors in two steps:  first find {$\ket{ABC} \approx a_0 U^{\dag} \ket{A B_L B_R C}$} for an initial guess of the isometry $a_0$ and unitary $U$ which minimizes the entanglement across the vertical bond highlighted in red;  second   set {$a = a_0 U^{\dag}$} and split $\ket{A B_L B_R C}$ in two via SVD.
The resulting $a$ comprise the tensors in $A^\ell$, and the choice of $U$ will produce a $\Lambda$ with  minimal vertical entanglement. 
     \label{fig:MM}} 
\end{figure}

It is actually sufficient to solve a simpler auxiliary problem: decompose $\Lambda^{\ell} = A^{\ell} \Lambda$, where $\Lambda$ is a wavefunction with only ancilla degrees of freedom (a ``zero-column'' wavefunction). The start and end points of the problem are shown in Fig.~\ref{fig:MM}a (i) and (v).
This move will be sufficient to move $\Lambda^\ell$ throughout the network, because we can tack the zero-column wavefunction onto the right in order to obtain the one-column wavefunction,  $\Lambda^{\ell+1} = \Lambda B^{\ell+1}$.

We can solve  $\Lambda^{\ell} \approx A^{\ell} \Lambda$ as a  variational problem, sweeping back and forth through the tensors to minimize $|\Lambda^{\ell} - A^{\ell} \Lambda | $ while respecting the isometry condition on $A$ and reducing the bond dimension of $\Lambda$ \cite{sup}.
Interestingly, however, we find a \emph{single} unzipping sweep based on disentangling provides a solution very close to the variational  one, \cite{sup} but is far quicker. This ``Moses Move'' (MM) is illustrated in sequence (i) to (v) of Fig.~\ref{fig:MM}a. 


The central  subproblem of the MM  (Fig.~\ref{fig:MM}b) takes in the orthogonality center $\ket{\lambda}$, which by grouping legs is a tripartite state  $\ket{ABC}$ on the top, lower left, and lower right degrees of freedom, and ``splits'' it into a four-partite state  $\ket{A B_L B_R C}$.
More precisely, we look for the splitting isometry $a^\dagger: B \to B_L \otimes B_R$, where  $a^\dagger a  = \mathds{1}$ and $B_{L/R}$ have dimension $\chi$, such that $\ket{A B_L B_R  C} = a^\dagger \ket{A B C}$ has minimal entanglement $S_{A B_L : B_R C}$.
This is closely related to finding the entanglement of purification\cite{terhal2002entanglement} of $\rho_{AC}$.
To do so we  make an initial (suboptimal) guess for the isometry $a_{0}$ chosen so that $B_L B_R$ includes the $\chi^2$  highest weight states in $B$, and then parameterize the optimal choice as {$a =  a_{0} U^{\dag}$} for a unitary $U$ acting on $B_L B_R$. We choose $U$ to minimize $S_{A B_L : B_R C}$ or it's Renyi generalization, \footnote{Results closest to the variational optimum are obtain if we use Renyi index $\alpha < 1$, see Supp.} a well defined optimization problem. \cite{Evenbly2014, hauschild2018finding}
The resulting $a$  comprise the isometries in $A^\ell$, allowing us to successively unzip $\Lambda^{\ell}$ into $A^{\ell} \Lambda$.

It would be  interesting to find a necessary and sufficient many-body entanglement criteria  for the success of the MM, in order to better understand why it finds a solution so close to the variationally optimal one.
In the absence of such rigorous results, we consider two practical numerical tests.

	
\begin{figure}
\centering
\includegraphics[width=0.95\columnwidth]{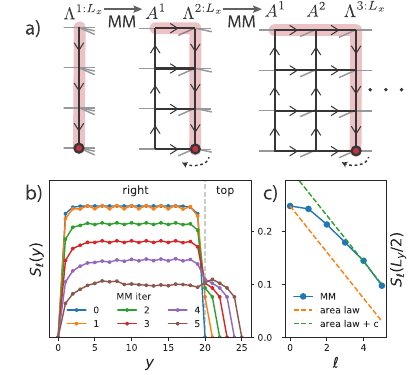}
\caption{The MPS to isoTNS algorithm: (a) The MPS $\Lambda^{1:L_x}$ for an $L_x \times L_y$ strip is fed into the MM by treating the legs of the first column as the  left ancilla and the remaining columns as the right ancilla to obtain $\Lambda^{1:L_x} = A^1 \Lambda^{2:L_x}$.
The renormalized wavefunction $\Lambda^{2:L_x}$ is then reshaped by viewing the legs of the second column as physical, and its vertical arrows are reversed downwards using the standard 1D MPS canonicalization algorithm.
Applying the MM again, we can repeat to obtain a canonical TNS. (b) Entanglement entropy $S_{\ell}$ for the sequence of orthogonality hypersurfaces (highlighted in pink) after $\ell$ iterations.
$y$ runs from bottom right, to top right, to top left. (c) $S_{\ell}$ for a cut at $y \sim L_y/2$, compared against the bulk area law determined from DMRG.
}\label{fig:colsplit}
\end{figure}
\emph{MPS to isoTNS}. Given a ground state wavefunction $\ket{\Psi}$ on an $L_x \times L_y$ strip, we propose an iterative algorithm to put $\ket{\Psi}$ into an isoTNS which we  test for the transverse Ising model $H=-\sum_{\langle i,j\rangle}\sigma^z_i\sigma_j^z - g\sum_i \sigma^x$ with Pauli matrices $\sigma^\mu$.
To implement it numerically, we consider a strip with $L_y \gg L_x$ and use DMRG to obtain the ground state as a 1D MPS $\Lambda^{1:L_x}$ where each ``site'' contains the $L_x$ spins of the corresponding row (Fig.~\ref{fig:colsplit}a).
As described in Fig.~\ref{fig:colsplit}, the MM can then be used to iteratively peal off columns of the wavefunction, $\Lambda^{\ell:L_x} = A^\ell \Lambda^{\ell+1:L_x} $, producing an isoTNS.
The algorithm is exponentially difficult in $L_x$ (since $\Psi$ is obtained as an MPS!), but serves as a check on the ansatz independent of a ground state search scheme.
Using an ancilla dimension $\chi = 6$ for the isometries, the  error $\| |\Psi_{\mathrm{MPS}}\rangle - |\Psi_{\mathrm{isoTNS}}\rangle \|^2$ is $2 \cdot 10^{-6}$ per site at $g=3.5$ (in the gapped, paramagnetic phase), $L_y = 20, L_x = 6,\chi_{\mathrm{MPS}}=128$, obtained in about 10 minutes on a laptop.

More interesting is the behavior of the ``vertical'' (top/bottom) and ``horizontal'' (left/right) entanglement of the resulting isoTNS.
At each step $\ell$ the orthogonality hypersurface $\Lambda^{\ell+1:L_x}$ makes a ``$\urcorner$'' shape, running up the right and over the top.
In Fig.~\ref{fig:colsplit}b we show the entanglement entropy $S_{\ell}(y)$ for cuts along $\Lambda$, and find $S_\ell$ decrease with $\ell$.
If the underlying phase has area law  $S_R =  s |\partial R | + \cdots$, for $y \sim L_y / 2$ we hope $S_{\ell}$  goes as $S_{\ell}  \approx  s (L_x - \ell) + \cdots $.
If not, the isometric columns $A^\ell$ aren't removing their share of the entanglement and the algorithm will fail in the thermodynamic limit.
In Fig.~\ref{fig:colsplit}c, we see that after the initial delay the algorithm  begins to remove remarkably close to $s$ entanglement per iteration.
The initial delay is expected, because  any two vertically-entangled degrees of freedom will  individually have some horizontal extent.
Until their entire support is to the left of $\Lambda$, the isometries $A$ cannot remove them.

The residual horizontal entanglement is left behind in the top-region of $\Lambda$.
As hoped for, the horizontal entanglement is of order $s$, and for $\ell = L_x$ we find $S_\ell$ smoothly matches up between the   right / top regions, despite the seemingly anisotropic nature of the algorithm.

\begin{figure}
\centering
\includegraphics[width=0.95\columnwidth]{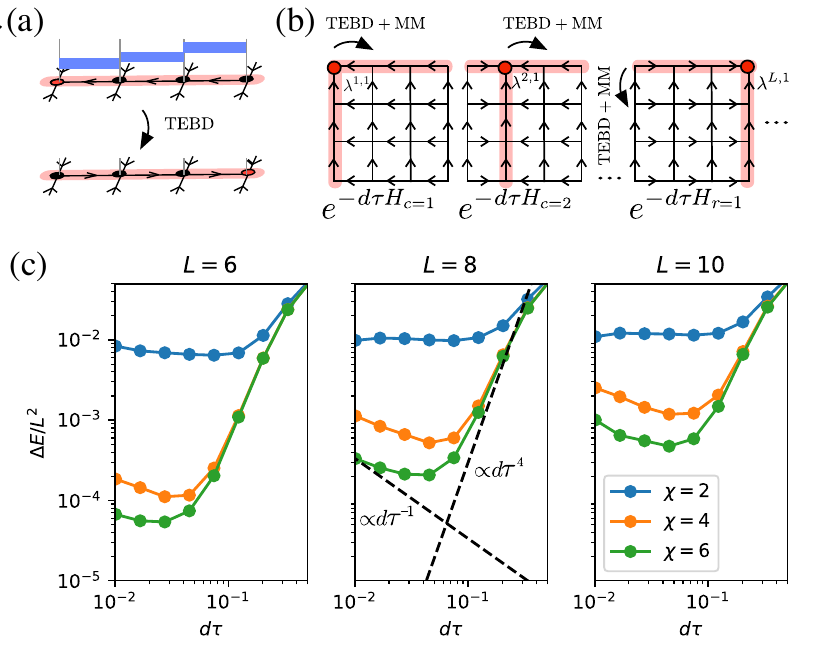}
\caption{The \tebdsq algorithm: (a) Trotterization of $e^{-\tau H_{r/c}}$ into a product of two-site terms acting on a single row/column of the isoTNS.   
(b) To complete one time step, the 1D update is applied to all row/columns by first applying the TEBD sweep and then sequentially shifting the orthogonality center $\lambda^{c,r}$ using the MM. Note that the update of one row/column reverses the arrows twice, e.g., the  TEBD sweep moves $\lambda^{c,r}$ from the top to the bottom and then MM moves it up again. 
(c) Error densities of the energy of the transverse field Ising model with $g=3.5$ for different system sizes and maximal bond dimensions $\chi$ as function of the Trotter step size $d\tau$.
}\label{fig:tebd}
\end{figure}

\emph{\tebdsq algorithm}. We now propose a Trotterized time stepper for isoTNS which can be used to obtain the ground state by imaginary-time evolution.
Assuming a nearest-neighbor interaction, we split the Hamiltonian into terms acting on columns and rows, $H = \sum^{L_x}_{c=1} H_c + \sum^{L_y}_{r=1} H_r$.
We Trotterize according to $e^{-\tau H} \approx  \prod_r e^{- \tau H_r}  \prod_c e^{- \tau H_c}$ as illustrated in Fig.~\ref{fig:tebd}a. As for the TEBD update in 1D, the  \tebdsq can be easily improved to second order. 
We start in canonical form with the orthogonality center $\lambda^{1,1}$ at site $c, r = 1,1$.
The evolution $e^{- \tau H_{c=1}}$ is then applied to column $\Lambda^1$ by calling the standard 1D TEBD algorithm \cite{Vidal2003a} at a cost $\propto\chi^6$.
We then use the MM to bring the orthogonality center over by one column, to $\lambda^{2,1}$ at a cost $\propto\chi^7$, and apply $H_{c=2}$, and so on, bringing the orthogonality center to $\lambda^{L_x, 1}$ (in contrast, the  full update of an unconstrained PEPS  costs $\chi^{12}$  \cite{verstraete:2004}).
Applying $e^{- \tau H_r}$ analogously  brings the center to $\lambda^{L_x, L_y}$, and we repeat to bring $\lambda$ counter-clockwise around the four corners to complete the  time step. 
Within a sweep the algorithm is  literally two nested versions of 1D TEBD (with the MM replacing QR/SVD  in the outer $c$-loop), hence the name ``\tebdsq''.

To benchmark \tebdsq, we return to the transverse field Ising model.
Fig.~\ref{fig:tebd}c  shows the energy density obtained from \tebdsq relative to  numerically exact results from large scale 1D-DMRG simulations at $g = 3.5$.
If the evolution were exact  the energy would decrease monotonically as the Trotter step $d \tau$ is decreased.
However, the MM has a small truncation error $\epsilon_{\textrm{MM}}$, and we see that the resulting energy has a minimum in $d \tau$.
For a $p$-th order Trotter step, the energy error should be $\Delta E =  a \, \epsilon_{\textrm{MM}} / d \tau + b \, d \tau^{2p}$ (in our implementation $p=2$), \footnote{Letting $\Delta$ be a typical energy scale and $\Delta E$ the energy density,  one step of imaginary time evolution decreases the energy density by $\Delta E \, \Delta \, d \tau$, while the Moses Move truncation increases it by $\epsilon_{MM} \Delta$. Thus at long times  the MM introduces $\Delta E \propto \epsilon_{\textrm{MM}} / d \tau$. } 
in agreement with the observed minima.
A similar effect is also observed in the full update of TNS, and can be partially remedied by using a variational update instead of imaginary time evolution \cite{Vanderstraeten2016,Corboz2016}.
The minimum energy converges towards the exact result as the bond dimension $\chi$ is increased.

\emph{Conclusions.} We introduced an isometric TNS ansatz which results in a canonical form that allows for 1D MPS algorithms to be efficiently adapted to 2D.
To numerically benchmark the ansatz,  we first demonstrated that an MPS representation of the ground state of the 2D transverse field Ising model can be efficiently transformed into an isoTNS.
Second, we implemented a \tebdsq algorithm and showed that it efficiently finds an approximation of the ground state of the 2D TFI model within the isoTNS form.
Future directions include theoretically understanding the variational power of the isoTNS ansatz, as well as implementing a variational ground state algorithm, DMRG$^2$, by nesting the standard 1D algorithm.


\acknowledgements We thank B. Bauer, M. Fishman, J. Haah, S. Lin, R. Mong, M. Stoudenmire, A. Turner, XL. Qi, and F. Verstraete for enlightening discussions.
We are indebted to R. Mong for pointing out the biblical origin of the Moses Move algorithm.
FP acknowledges the support of the Deutsche Forschungsgemeinschaft (DFG, German Research Foundation) Research Unit FOR 1807 through grants no. PO 1370/2-1, TRR80, Germany's Excellence Strategy -- EXC-2111 -- 390814868, and the European Research Council (ERC) under the European Union as Horizon 2020 research and innovation program (grant agreement no. 771537).
MZ was funded by the U.S. Department of Energy, Office of Science, Office of Basic Energy Sciences, Materials Sciences and Engineering Division under Contract No. DE-AC02-05-CH11231 through the Scientific Discovery through Advanced Computing (SciDAC) program (KC23DAC Topological and Correlated Matter via Tensor Networks and Quantum Monte Carlo).  
This work was finished at the Aspen Center for Physics, which is supported by National Science Foundation grant PHY-1607611.

\bibliography{biblio}
\pagebreak
\onecolumngrid
\appendix

\section*{Supplemental Material}

\section{Variational solution of $\Lambda^{\ell} = A^\ell \Lambda$}
The Moses Move is not strictly variational: it sweeps only once through the network and chooses $A^\ell$ according to an entanglement criteria, rather than maximization of the global overlap.
Here we describe a complementary variational procedure which can be used to further optimize the decomposition after the MM, though in practice we find the MM is surprisingly close to optimal (see below).

We optimize the overlap $\braket{A^\ell \Lambda | \Lambda^{\ell}  }$ by a variant of alternating least squares, e.g., optimize the tensors one at a time, subject to the constraints.
We do so by  first fixing $A^\ell$ and viewing the problem as maximization of $\textrm{max}_{\Lambda}  \braket{ \Lambda | {A^\ell}^\dagger \Lambda^{\ell}  } $ over the zero-column MPS $\Lambda$.
This is the completely standard problem of variational MPS compression which is described elsewhere \cite{verstraete2008matrix,schollwoeck:2011}.
We then hold $\Lambda$ fixed and sweep up through the network to optimize each $A^\ell_y$ on site $y$.
Here the isometric constraint comes in, so we use the polar-decomposition algorithm for optimizing over isometries \cite{Evenbly2014}.
We consider the variation of the overlap with respect to $A^\ell_y$, and reorganize the SVD decomposition of the variation as follows:
\begin{align}
 \frac{d}{d {A^\ell_y}^\dagger} \braket{ A^\ell \Lambda  | \Lambda^{\ell}  } =  U s V = (U V) (V^\dagger s V)
\end{align}
To decode the index structure, we first view $A^\ell_y$ as a rank-two matrix by grouping together the incoming / outgoing arrows, so the variation is a matrix we can SVD.
The isometry is then updated according to $A^\ell_y \to U V$. Using the MM as an initial guess, we find a slow but monotonic convergence to an optimal solution. 

\section{Comparison of Moses Move with variationally optimal solution}
Here we compare the Moses Move (MM) solution of $\Lambda^{\ell} = A^\ell \Lambda$ with the variational method just discussed. 
The difference varies from case by case, so we consider some representative data for the ground state $\Psi$ of the transverse field Ising model on a $2 \times 20$ site ladder obtained from DMRG. The basic structure of the MM is to factorize a wavefunction with some ``left'' degrees of freedom $i$ and right degrees of freedom $j$ according to $\Psi_{ij} = \sum_k A_{i k} \Lambda_{k j}$ where $A$ is an isometry, and $A, \Lambda$ have the usual MPO structure. Here left / right will correspond to the first / second column of $\Psi$. We choose the isometric MPO $A$ to have dimensions $D_H = D_V = 2$ in the bulk of the strip, while for $\Lambda$, $D_H = 2, D_V = \eta$, and we'll measure how $| \Psi - A \Lambda |$ depends on $\eta$.

\begin{figure}
\centering
\includegraphics[width=0.5\columnwidth]{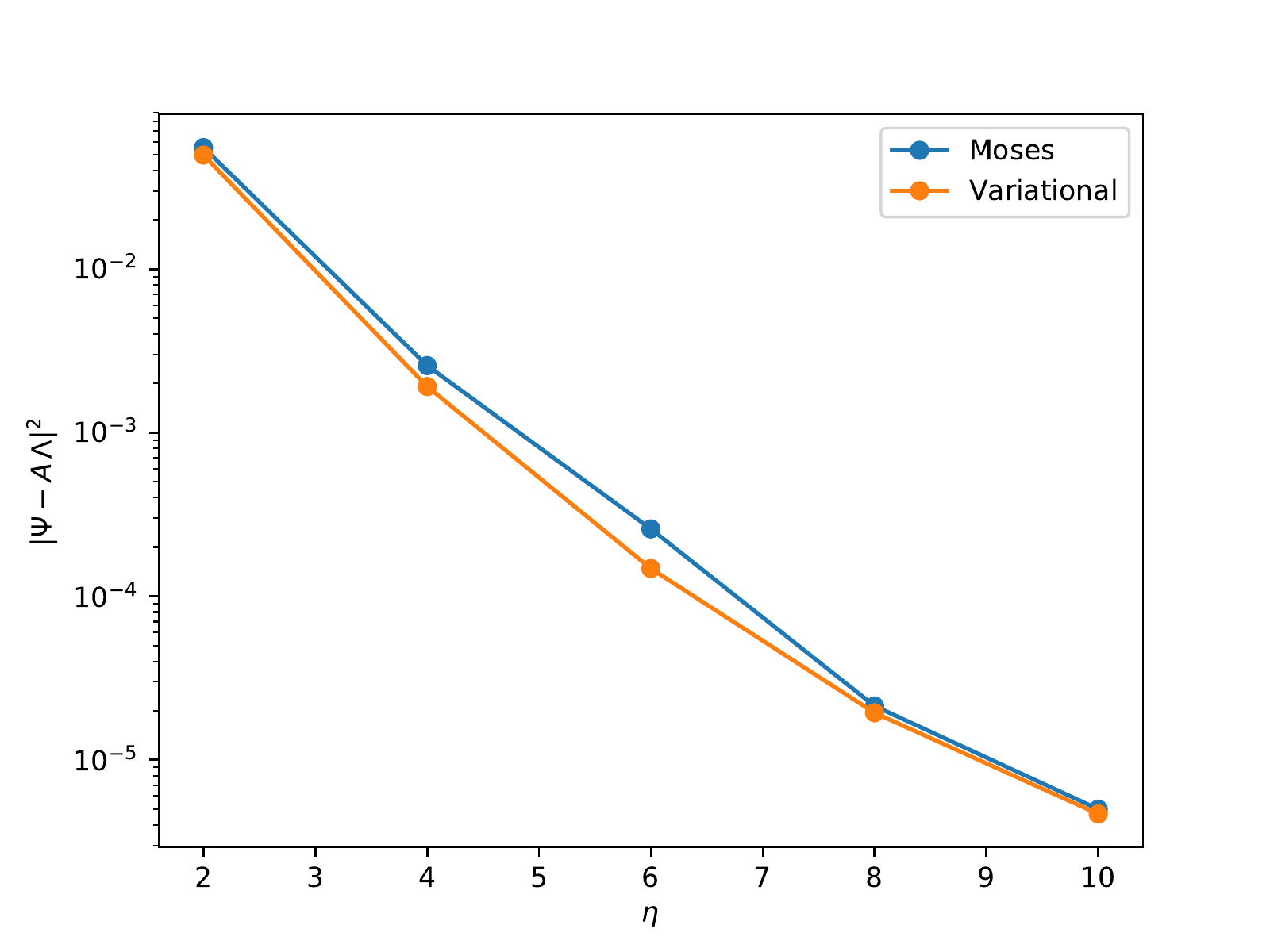}
\caption{Error in the decomposition $| \Psi - A \Lambda |^2$ for both the Moses Move (with disentangling based on $\alpha = \frac{1}{2}$ Renyi entropy) and the variationally optimized solution. $\eta$ is the vertical bond dimension of $\Lambda$. }\label{fig:var_vs_mm}
\end{figure}

In the MM, recall the tensor $T$ which splits $T \ket{ABC} = \ket{AB_L: B_R C}$ is chosen to minimize the entanglement $S_{A B_L: B_R C}$. A priori, there is no reason we need to use the von-Neumann entropy for our measure, and in fact, because we want to minimize the bond dimension $\eta$ of $\Lambda$, it is better to use a Renyi entropy $S^{[\alpha]}$ with $\alpha < 1$.
It is known that $S^{[\alpha]}$ bounds the accuracy of a matrix product representation for any $\alpha < 1$, while $\alpha \geq 1$ does not. \cite{verstraete2006matrix}
Indeed, we find that the difference between the MM and variational error decreases over the range $\frac{1}{2} \lesssim \alpha \lesssim 2$, but (for the cases we have examined) there isn't a significant difference beyond this. So here we use $\alpha = \frac{1}{2}$. The resulting error is shown in Fig.~\ref{fig:var_vs_mm}. We compare it with a factorization in which the MM is further optimized by the previously discussed variational sweeps, which presumably finds the optimal solution modulo issues of local minima.
The difference between the MM error and the optimal one is below a factor of 2 for all  $\eta$. Practically speaking, this is a very small difference because the error is reduced by an \emph{order of magnitude} just by increasing $\eta$ slightly. So (we believe) it is  more efficient to  use the MM, which is drastically faster, and increase  $\eta$ slightly, rather than obtain the variationally optimal solution at each step. 

\section{Index structure of the ``splitting'' step of the MM}
Here we make explicit the index structure of the central ``splitting'' step of the MM. By grouping legs, the central site wavefunction $\lambda$ is viewed a tripartite state $\lambda_{\alpha \beta \gamma}$ where indices $\alpha, \beta, \gamma$ run over parties $A, B, C$. $\lambda$ is then split into the product of three tensors $a, \lambda', V$ as follows:
\begin{align}
\lambda_{\alpha \beta \gamma} &\approx \sum_{\beta_L, \beta_R} a_{\beta, \beta_L \beta_R} \Theta_{\alpha \beta_L \beta_R \gamma},  \quad   a^\dagger a = \mathds{1}_{B_L B_R}\\
 \Theta_{\alpha \beta_L \beta_R \gamma} &= \sum_k (U_{\alpha \beta_L, k} s_k) V_{k, \beta_R \gamma} \approx \sum^\chi_k \lambda'_{\alpha \beta_L, k} V_{k, \beta_R \gamma}
\end{align}
Given a choice for the isometry $a^\dagger: B \to B_L B_R$, we define the state $\Theta$ on $A B_L B_R C$ by $\Theta = a^\dagger \lambda$. 
The error at this step is $| \lambda - a \Theta|^2 = | (1 - a a^\dagger ) \lambda|^2$. This is just the truncated weight from the projection $a^\dagger : B \to B_L B_R$.  Next, we bipartition according to $(A B_L) (B_R C)$ and Schmidt decompose $\Theta = U s V$. The tensor $\lambda' = U s$ is then defined by truncating the Schmidt spectrum to the $\chi$-most important states.
In order to maximize the fidelity, the splitting isometry $a$ is chosen to minimize an $\alpha$-Renyi entropy $S^{[\alpha]}$ of the Schmidt spectrum $s_k$ ($\alpha = \frac{1}{2}$ gives good results).
The tensor $a$ becomes the constituent tensor of $A^\ell$, $V$ is the constituent tensor of $\Lambda$, while $\lambda'$ is the new orthogonality center which gets merged upwards into the next step of the MM.

\end{document}